\documentclass[12pt]{iopart}  
\usepackage{amssymb}
\usepackage{bbm,graphicx}   

\def\ave#1{\langle #1 \rangle}
\def\ii{{\rm i}}
\def\ket#1{\vline\, #1 \rangle}
\def\bra#1{\langle #1\, \vline}
\def\braket#1#2{\langle #1\, \vline\, #2 \rangle}

\def\tit#1{{\em #1},}
\def\etal#1{#1}

\begin{document}

\title{Transport in a one-dimensional isotropic Heisenberg model at high temperature}

\author{Marko \v Znidari\v c}
\address{
Instituto de Ciencias F\' isicas, Universidad Nacional Aut\' onoma de M\' exico, Cuernavaca, Mexico, and\\
Faculty of Mathematics and Physics, University of Ljubljana, Ljubljana, Slovenia}

\date{\today}

\begin{abstract}
Magnetization transport in a one-dimensional isotropic spin $1/2$ Heisenberg model is studied. It is shown that in a nonequilibrium steady state at high temperature and constant small driving the magnetization current depends on the system length $L$ as $\sim 1/L^{0.5}$, meaning that the diffusion constant diverges as $\sim L^{0.5}$. Spectral properties of a superoperator governing the relaxation towards a nonequilibrium steady state are also discussed.
\end{abstract}
%\submitto{{\it J.~Stat.~Mech.}}
%05.30.-d       Quantum statistical mechanics
%03.65.Yz 	Decoherence; open systems; quantum statistical methods
%05.70.Ln 	Nonequilibrium and irreversible thermodynamics
%75.10.Pq 	Spin chain models       
%05.60.Gg 	Quantum transport

\section{Introduction}

The Heisenberg model of nearest-neighbor coupled spins is of high interest in theoretical as well as in experimental physics. The simplest variant is a one-dimensional (1d) chain of coupled spin-1/2 particles, described by the Hamiltonian
\begin{equation}
H= \sum_{j=1}^{L-1} \sigma^{\rm x}_j  \sigma_{j+1}^{\rm x} +\sigma^{\rm y}_j  \sigma_{j+1}^{\rm y}  + \sigma^{\rm z}_j  \sigma_{j+1}^{\rm z},
\label{eq:H}
\end{equation}
in terms of Pauli matrices $\sigma^{\rm x,y,z}_j$ at lattice site $j$. Using Jordan-Wigner transformation~\cite{JW} it can be mapped to a system of spinless fermions whose Hamiltonian has a kinetic (hopping) term and a density-density interaction term between fermions at neighboring sites. It therefore represents one of the simplest systems of strongly interacting fermions. The model (\ref{eq:H}) is exactly solvable by the Bethe ansatz~\cite{Bethe}. Despite its solvability, time dependent properties, like a time-dependent current autocorrelation function that is via a linear response theory directly related to the transport coefficient, are at present beyond capabilities of exact methods. Isotropic one-dimensional Heisenberg model (\ref{eq:H}) is experimentally realized in so-called spin-chain materials~\cite{spin-chain}, for instance in many cuprates. As of yet unexplained in these materials is a very high thermal conductivity along the axis of Heisenberg chains~\cite{exper}, believed to be due to contribution from Heisenberg chains and strongly influenced by impurities.

In the present work we shall study magnetization transport in the isotropic Heisenberg chain; for an overview of references on a more general anisotropic Heisenberg model see introductions in Refs.~\cite{pregled,Sirker11}. Isotropic Heisenberg model has been studied in the past, however, no definite conclusion about magnetization transport has been reached so far. Most studies focused on the Drude weight, whose non-zero value indicates a non-diffusive transport, usually just called ballistic. Using the Bethe ansatz a non-zero Drude weight at all temperatures is advocated in~\cite{Benz}, with $\sim 1/T$ behavior at high temperatures, while zero Drude weight at all temperatures is predicted in~\cite{Zotos}. Exact diagonalizations~\cite{Heidrich:03}, as well as conformal field theory~\cite{Fujimoto} and quantum Monte Carlo~\cite{Sorella}, also result in a finite Drude weight at high temperatures, see also~\cite{Narozhny,Mukerjee,Herbrych}. Using current autocorrelation function obtained by exact diagonalization is non-conclusive~\cite{Steinigeweg} because long time scales exist. Extrapolation to the thermodynamical limit is very difficult with these results because exact diagonalization is limited to small systems, while quantum Monte Carlo, Bethe ansatz and conformal approaches have their own problems~\cite{Sirker11}. Low-energy bosonization calculation, together with the analysis of current autocorrelation function from density matrix renormalization group method, on the other hand indicates~\cite{Sirker11} a presence of diffusive contribution, also seen in quantum Monte Carlo calculation~\cite{Grossjohann}. All these results, some showing zero Drude weight, others non-zero, or even the presence of a diffusive component, call for a more precise characterization of transport. Quantification just in terms of zero or non-zero Drude weight namely fails to distinguish between different variants of non-diffusive behavior. 

A more detailed classification can be done for instance by studying how the current $j$ scales with system length $L$ if one fixes the difference of potentials at boundaries. Two extreme examples are the scaling $j \sim 1/L$ if a system obeys Fourier law, and $j \sim L^0$ in case of ballistic transport. However, as is well known from studies of classical transport~\cite{review-class}, in many systems the scaling exponent is a real number, $j \sim L^{-\alpha}$, with $0 \le \alpha \le \infty$. Transport is called anomalous if the scaling exponent differs from $1$, $\alpha\neq 1$. The exponent $\alpha$ is via the Fourier-like law that relates a transported quantity $z$ and its current $j$, $j = -D\, \nabla z$, directly related to the diffusion constant $D$, resulting in the scaling $D \sim L^{1-\alpha}$. Under certain assumptions a heuristic argument with classical non-interacting particles leads to a connection between $\alpha$ and the exponent $\beta$ of the spreading of disturbances as quantified by the variance $\sigma^2$, $\sigma^2 \sim t^\beta$. The relation is $\beta=\frac{2}{1+\alpha}$~\cite{Li}. The regime of $1 < \beta \le 2$, corresponding to $0 \le \alpha <1$, is called super-diffusion, while that of $0 \le \beta < 1$, corresponding to $\alpha>1$, is called sub-diffusion. Note that sub-ballistic transport with $\beta < 2$ means that the Drude weight is zero.

Recently, numerical simulations have shown~\cite{PRL11} that $\alpha=0.5$ in the isotropic Heisenberg model at an infinite temperature in the linear response regime. In the present work we shall extend on these results by calculating diffusion constant also at finite temperatures, showing that the scaling stays the same at high temperatures (higher than the exchange interaction). We shall also provide some other properties of isotropic Heisenberg model, like the relaxation rate to a nonequilibrium steady state.

\section{The Method}
In order to be able to study nonequilibrium stationary states (NESS) we couple the system to reservoirs at left and right chain ends. The two reservoirs are kept at different potentials inducing a nonzero magnetization current through the chain. We describe reservoirs in an effective way using the Lindblad equation~\cite{Lin} for the density matrix $\rho$ of the system,
\begin{equation}
{{\rm d}}\rho/{{\rm d}t}=\ii [ \rho,H ]+ {\cal L}^{\rm dis}(\rho)={\cal L}(\rho),
\label{eq:Lin}
\end{equation}
where the dissipative linear operator ${\cal L}^{\rm dis}$ describing bath is expressed in terms of Lindblad operators $L_k$, ${\cal L}^{\rm dis}(\rho)=\sum_k \left( [ L_k \rho,L_k^\dagger ]+[ L_k,\rho L_k^{\dagger} ] \right)$.

We use two kinds of reservoirs. To obtain NESS at infinite temperature, i.e., zero energy density, we use the so-called one-spin bath which is realized by two Lindblad operators at each end, $L^{\rm L}_1=\sqrt{1-\mu}\,\sigma^+_1, L^{\rm L}_2= \sqrt{1+\mu}\, \sigma^-_1$ at the left end and $L^{\rm R}_1 =  \sqrt{1+\mu}\,\sigma^+_n, L^{\rm R}_2= \sqrt{1-\mu}\, \sigma^-_n$ at the right end, $\sigma^\pm=(\sigma^{\rm x} \pm {\rm i}\, \sigma^{\rm y})/2$, with $\sigma^{\rm x,y,z}$ being Pauli matrices. 

To simulate NESS at a finite temperature we use the so-called two-spin bath, in which one has 16 Lindblad operators acting at two boundary spins at each end. The form of these 16 operators is complicated and we do not state it explicitly. They are obtained by demanding that the stationary equation on two boundary spins, described by $\rho^{(2)}$, ${\cal L}^{\rm dis}(\rho^{(2)})=0$, has for a solution a grandcanonical state $\rho^{(2)}$. Targeted grandcanonical state $\rho^{(2)}$ is obtained by calculating it from a small chain of $8$ spins, $\rho^{(2)} \sim \tr_{3,\ldots,8}( \exp{(-H/T_{\rm L,R}+\mu_{\rm L,R} M)}) $ (tracing is performed over $6$ spins in a chain of length $8$), where $M=\sum_{j=1}^L \sigma_j^{\rm z}$ is a total magnetization and $T_{\rm L,R}$ and $\mu_{\rm L,R}$ the imposed temperature and potential at the left/right end of the chain. Details of the implementation can be found in~\cite{JSTAT09}. Note that $\rho^{(2)}$ is used only to generate appropriate Lindblad operators that will simulate finite temperature. Once a NESS is found for a large chain (of upto 256 spins) a real system's temperature will be determined by calculating expectation values of observables in the NESS. Our results do not depend on details of Lindblad operators used in the simulation (therefore also not on details of $\rho^{(2)}$ used in deriving them), their only goal is to impose a finite energy density in the NESS. The driving parameter $\mu_{\rm L,R}$ (or $\mu$) will always be small as we are interested in the linear response regime. For stationary properties under maximally strong one-spin driving see~\cite{strong}. 

For both kinds of reservoirs the NESS, simply denoted by $\rho$ in the rest of the paper, is found by evolving an arbitrary initial state $\rho(0)$ with the Lindblad equation for sufficiently long time, until a nonequilibrium stationary state $\rho=\lim_{t \to \infty}{\rho(t)}$ is reached. To calculate time evolution of $\rho(t)$ we use a time-dependent density matrix renormalization group (tDMRG) method, as described in refs.~\cite{JSTAT09}.

In section~\ref{sec:33} we will be interested also in spectral properties of the Liouville superoperator ${\cal L}$ (\ref{eq:Lin}). Because a detail knowledge of the eigenvalues of ${\cal L}$ can not be obtained by a simple implementation of tDMRG we use, we have instead used an exact diagonalization on somewhat smaller systems.

\section{Results}

\subsection{Determining temperature}
In all our simulations, using one-spin or two-spin baths, we use a weak driving $\mu_{\rm L}=-\mu=-0.02$ and $\mu_{\rm R}=+\mu=0.02$. Symmetric driving with respect to left/right end imposes a NESS state with the average magnetization being zero. In a two-spin bath, with which we can simulate finite temperature states, the imposed temperature is the same at the left and the right end, $T_{\rm L}=T_{\rm R}=T_{\rm imp.}$. A consequence of this is that the energy density in the NESS,
\begin{equation}
h_j=\ave{\sigma^{\rm x}_j  \sigma_{j+1}^{\rm x} +\sigma^{\rm y}_j  \sigma_{j+1}^{\rm y}  + \sigma^{\rm z}_j  \sigma_{j+1}^{\rm z}},
\label{eq:hj}
\end{equation}
is independent of the position index $j$ and the energy current in the NESS is therefore zero. $\ave{}$ denotes the expectation value in the NESS, $\ave{A}=\tr{\rho\, A}$. Because driving is weak the NESS is locally close to equilibrium. We can therefore determine~\cite{thermal} the temperature of NESS, called a ``measured'' temperature $T_{\rm meas.}$, by equating the expectation value of the energy density to the one expected in a canonical state, $h_j=h_{\rm T}$,
\begin{equation}
h_{\rm T}=\frac{\tr{(\rho_{\rm T} H)}}{L-1},\qquad \rho_{\rm T}=\frac{\exp{(-H/T_{\rm meas.})}}{\tr{\,\exp{(-H/T_{\rm meas.})}}},
\label{eq:ET}
\end{equation}
and solving for $T_{\rm meas.}$. The NESS can be approximated by a canonical state because the average potential $\mu_{\rm meas.}$ is zero. We have checked that the NESS state is indeed close to the canonical one by verifying that expectation values of nearest-neighbor observables in the bulk of the NESS agree with the canonical ones within ${\cal O}(\mu_{\rm L,R})$. For instance, for the NESS with $T_{\rm meas.}\approx 4.8$ we have one-point expectation values $\ave{\sigma_k^{\rm x,y}} < 10^{-5}$ (they are non-zero due to tDMRG truncation error), $\ave{\sigma_k^{\rm z}} \sim {\cal O}(\mu_{\rm L,R})$, two-point expectation values $\ave{\sigma_k^{\rm x} \sigma_{k+1}^{\rm x}}\approx \ave{\sigma_k^{\rm y} \sigma_{k+1}^{\rm y}} \approx \ave{\sigma_k^{\rm z} \sigma_{k+1}^{\rm z}} \approx h_{\rm T}/3$, $\ave{j_k} \sim {\cal O}(\mu_{\rm L,R})$, while other two-point nearest-neighbor expectation values are smaller than $10^{-5}$. In the thermal canonical state all one-point expectation values are zero, while two-point nearest neighbor agree with the ones in the NESS. Three-point expectation values in the NESS are all smaller than $10^{-5}$, except $\ave{\sigma_k^{\rm x} \sigma_{k+1}^{\rm x}\sigma_{k+2}^{\rm z}}$ and $\ave{\sigma_k^{\rm y} \sigma_{k+1}^{\rm y}\sigma_{k+2}^{\rm z}}$ (as well as that of all permutations of the three Pauli matrices occurring in these two operators) which are of order $10^{-4}$. In the canonical state all three-point nearest-neighbor expectation values are zero. Some four-point nearest neighbor expectation values in the NESS are non-zero and of size $\sim 0.05$, for instance of operators like $\sigma_k^{\rm x} \sigma_{k+1}^{\rm x}\sigma_{k+2}^{\rm x}\sigma_{k+3}^{\rm x}$, $\sigma_k^{\rm x} \sigma_{k+1}^{\rm x}\sigma_{k+2}^{\rm z}\sigma_{k+3}^{\rm z}$ or $\sigma_k^{\rm x} \sigma_{k+1}^{\rm x}\sigma_{k+2}^{\rm y}\sigma_{k+3}^{\rm y}$ (and permutations of these four operators). The corresponding canonical expectation values are also non-zero and within ${\cal O}(\mu_{\rm L,R})$ of the NESS expectation values. All canonical expectation values mentioned have been calculated at high temperatures by an exact diagonalization of small systems while the imaginary-time tDMRG method has been used at smaller temperatures.

\begin{figure}
\centerline{\includegraphics[width=0.7\textwidth]{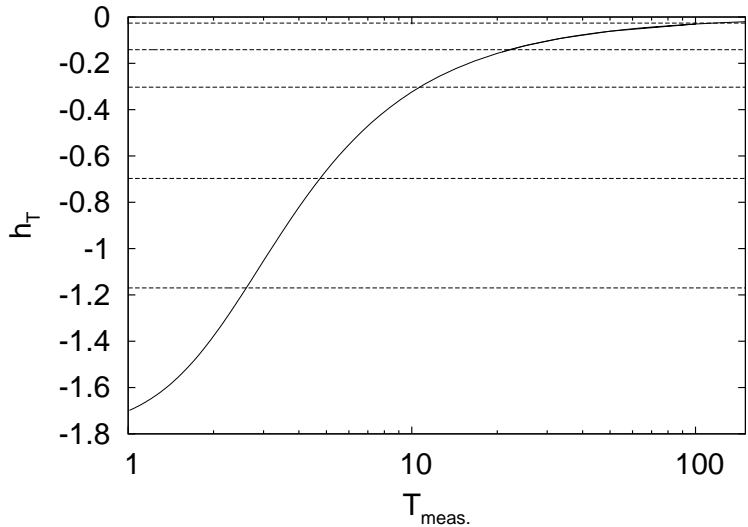}}
\caption{Average energy density in a canonical state (\ref{eq:ET}) of the isotropic one-dimensional Heisenberg model (\ref{eq:H}). Horizontal lines show energy densities $h_j$ in the NESS states used in the paper.}
\label{fig:kanon}
\end{figure}

In Fig.~\ref{fig:kanon} we show the dependence of the canonical energy density on temperature. 
It turns out that the boundary effects with our two-spin bath (see later) get increasingly stronger with lowering the imposed temperature. Because the operator entanglement of the NESS $\rho$ also increases, simulations get increasingly more difficult. We are therefore not able to reach very low temperatures. The temperatures we used are listed in Table~\ref{tab:T}. 
\begin{table}[ht!]
\begin{center}
\begin{tabular}{r||c|c}
$T_{\rm imp.}$ & $h_j$ & $T_{\rm meas.}$ \\
\hline
$\infty$ & 0.0004 & $\infty$ \\
50 & -0.026 & 120 \\
10 & -0.141 & 22 \\
5 & -0.30 & 10.6 \\
2 & -0.69 & 4.8 \\
0.2 & -1.17 & 2.6
\end{tabular}
\end{center}
\caption{Data for NESS states used in the paper. For $T_{\rm imp.}=\infty$ we use a one-spin bath, for others a two-spin bath. The measured temperature is determined by equating energy density $h_j$ in the NESS to the canonical expectation value $h_{\rm T}$. A nonzero $h_j$ for $T_{\rm imp.}=\infty$ is due to truncation errors of the tDMRG method.}
\label{tab:T}
\end{table}
Note that the minimal temperature achieved, $T_{\rm meas.}=2.6$, would in the spin notation where $H=\sum_{j=1}^{L-1} s^{\rm x}_j  s_{j+1}^{\rm x} +s^{\rm y}_j  s_{j+1}^{\rm y}  + s^{\rm z}_j  s_{j+1}^{\rm z}$, with $s^{\rm x,y,z}=\sigma^{\rm x,y,z}/2$, correspond to $T_{\rm meas.}=0.65$. The exchange interaction in ${\rm SrCuO}_2$ is approximately $J/k_{\rm B} \approx 2000$ K. Temperatures in the experiments~\cite{exper}, which are of the order $\sim 100$ K, therefore correspond to dimensionless temperature $T_{\rm meas.} \approx 0.2$ in our Pauli notation. Such low temperatures are unfortunately not reachable with our reservoirs~\cite{thermal}.

\subsection{Magnetization profiles and the current}
\begin{figure}[ht!]
\centerline{\includegraphics[width=0.7\textwidth]{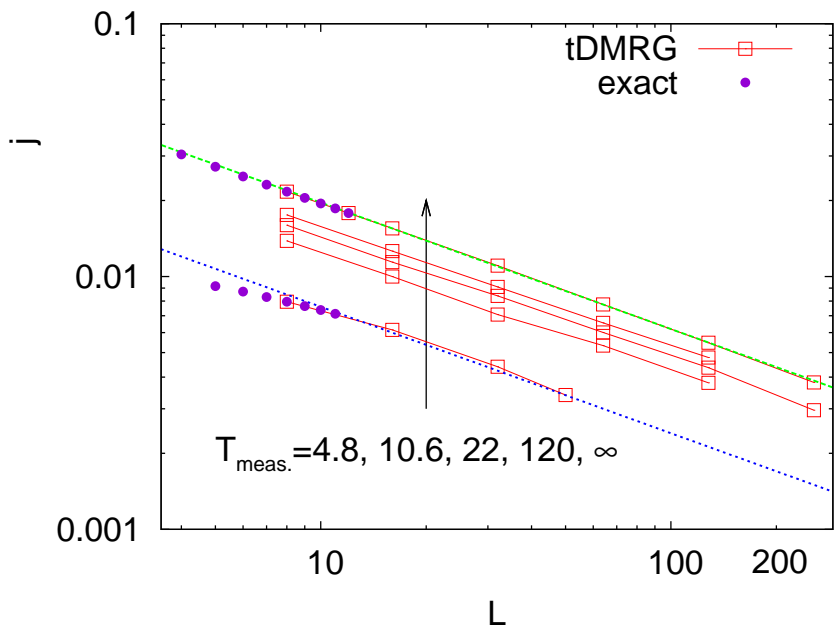}}
\caption{Scaling of current $j$ with $L$ for different temperatures $T_{\rm meas.}$, all at the same driving $\mu_{\rm L,R}$. Circles are using numerically exact NESS state while squares are obtained using tDMRG. Data at $T_{\rm meas.}=\infty$ is the same as in Ref.~\cite{PRL11}, apart from new data point for $L=256$. Two straight lines overlapping with $T_{\rm meas.}=4.8$ and $T_{\rm meas.}=\infty$ data are $\sim 1/L^{0.5}$.
}
\label{fig:j_odn}
\end{figure}

\begin{figure}[ht!]
\centerline{
\includegraphics[width=0.5\textwidth]{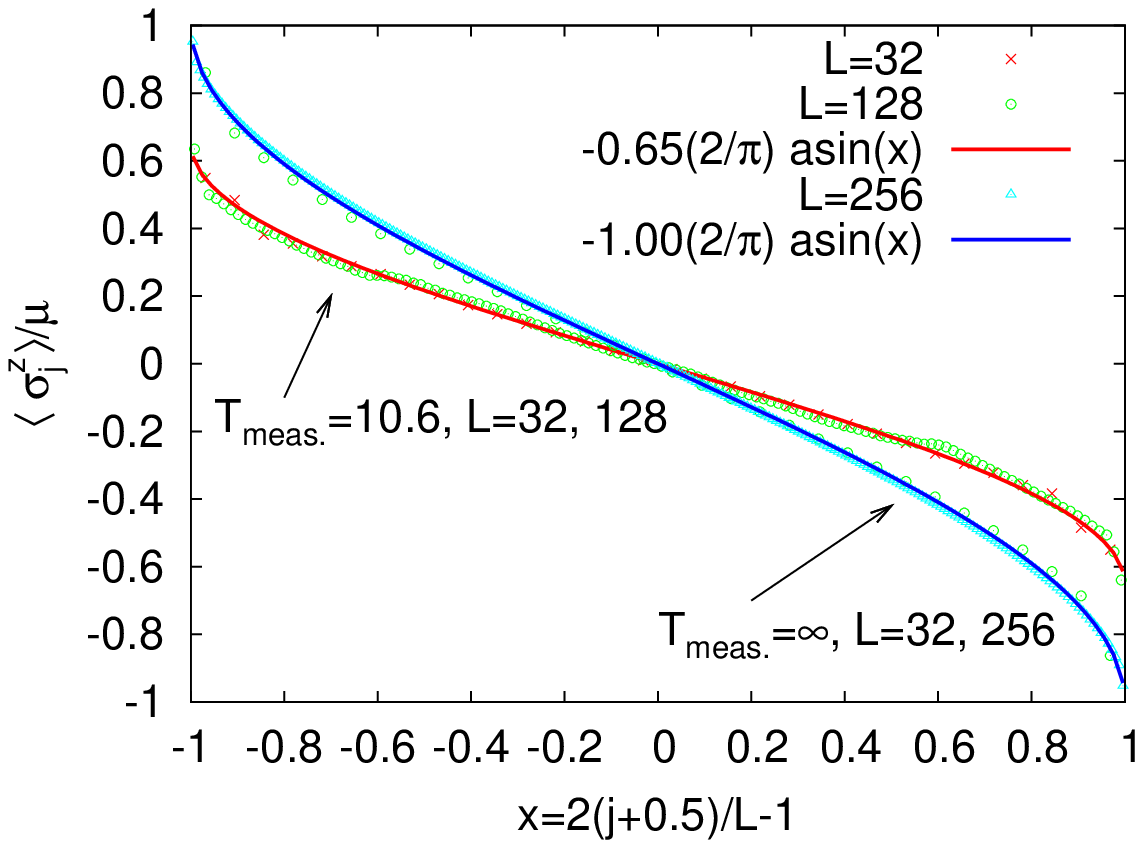}
\includegraphics[width=0.5\textwidth]{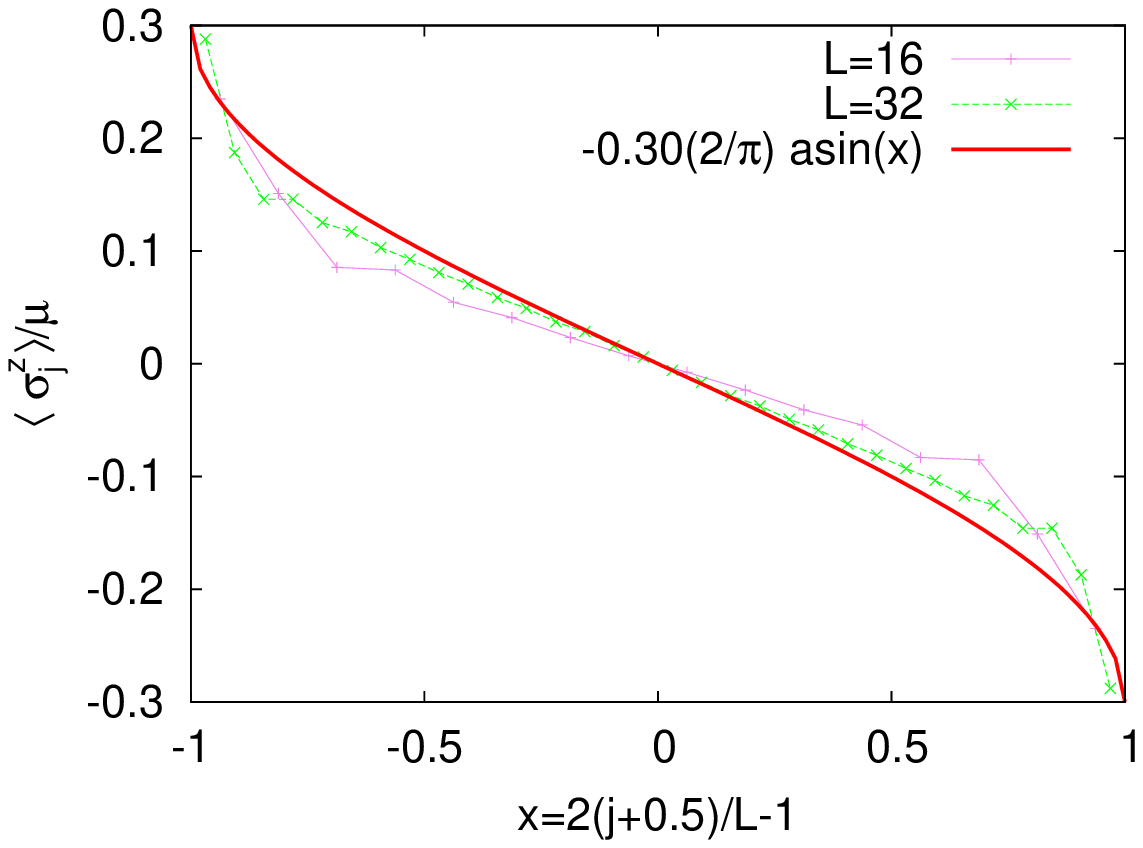}
}
\caption{(Color online) Left: Magnetization profiles at $T_{\rm meas.}=\infty$ and at $T_{\rm meas.}=10.6$. At lower temperature the magnetization exhibits jumps at the boundary. At all temperatures the profile is well described by the scaling function $\arcsin{(x)}$. Right: At lower $T_{\rm meas.}=4.8$ the convergence of profiles to $\arcsin{(x)}$ seems to happen at larger sizes $L$ than at higher temperatures.}
\label{fig:zi}
\end{figure}

\begin{figure}[ht!]
\centerline{\includegraphics[width=0.7\textwidth]{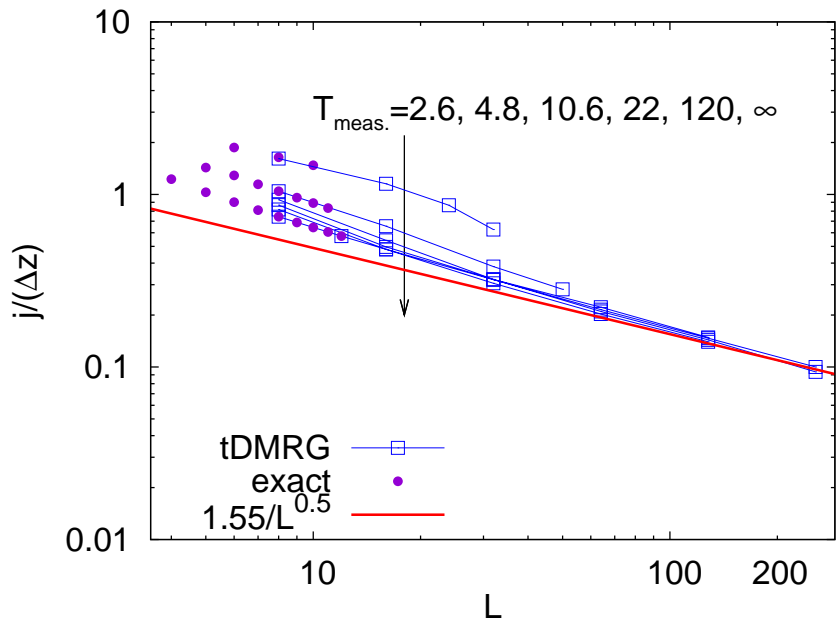}}
\caption{Current divided by $\Delta z=\ave{\sigma_1^{\rm z}-\sigma_L^{\rm z}}$ for different $L$ and temperatures. For sufficiently large $L$ all seem to converge to the same line $\sim 1.55/L^{0.5}$.}
\label{fig:jr}
\end{figure}

The main quantity we consider is the scaling of the expectation value of the magnetization current in the NESS with the system size $L$. The magnetization current operator is,
\begin{equation}
j_k=2(\sigma^{\rm x}_k \sigma^{\rm y}_{k+1}-\sigma^{\rm y}_k \sigma^{\rm x}_{k+1}),
\label{eq:jk}
\end{equation}
and we denote its expectation value~\cite{footnote} in the NESS, which is independent of the site index $k$, simply by $j=\ave{j_k}$. In Fig.~\ref{fig:j_odn} we show results at various temperatures, all for the same driving $\mu_{\rm L,R}=\pm 0.02$. In addition to NESS results obtained by tDMRG we also show the ones obtained by numerically exactly solving~\cite{arpack} the master equation (\ref{eq:Lin}).
The reason that in general the current $j$ decreases with decreasing temperature is also a consequence of increasing boundary resistances due to two-spin reservoirs used. There is a magnetization jump at the boundary so that the first and the last spin have magnetization smaller than the imposed $\mu=0.02$. This can be seen in Fig.~\ref{fig:zi} for data at finite temperatures.
The magnetization profiles at all temperatures from Table~\ref{tab:T} (except the ones at $T_{\rm meas}=2.6$) can be well described by the scaling function $\ave{\sigma_r^{\rm z}}= k\frac{2\mu}{\pi} \arcsin(x)$, where $x=2\frac{r-0.5}{L}-1$ is a scaled position and $k$ is a temperature dependent prefactor, effectively taking into account for boundary jumps. For instance, at $T_{\rm meas.}=10.6$ (left frame in Fig.~\ref{fig:zi}) it is $k \approx 0.65$. Because of the boundary jumps, to properly account for the scaling of $j$ with $L$, ie., to asses the validity of the Fourier law, 
\begin{equation}
j=-D\, \nabla_r \ave{\sigma_r^{\rm z}},
\label{eq:Four}
\end{equation}
we have to scale the current with the actual magnetization difference given by $\ave{\sigma_1^{\rm z}}-\ave{\sigma_L^{\rm z}}\equiv \Delta z$. This is shown in Fig.~\ref{fig:jr}.
From the figure we can estimate that the current is
\begin{equation}
j = -1.55 \frac{\ave{\sigma_1^{\rm z}}-\ave{\sigma_L^{\rm z}}}{L^{0.5}},
\label{eq:cur155}
\end{equation}
independent of the temperature (energy density), at least for $T_{\rm meas.} \gtrsim 5$. In this range of temperatures the diffusion constant $D$ (\ref{eq:Four}) therefore scales as
\begin{equation}
D \approx 1.55\, L^{0.5}.
\label{eq:D}
\end{equation}
At lower temperatures it is difficult to assess if the scaling is still the same. Data in Fig.~\ref{fig:jr} for $T_{\rm meas.}=2.6$ and $4.8$ show larger current than predicted by Eq.(\ref{eq:cur155}), however, one explanation could be that the length at which the asymptotic behavior (\ref{eq:cur155}) begins gets larger than the sizes studied. In the right frame of Fig.~\ref{fig:zi} we can for instance see that the asymptotic $\arcsin{x}$ magnetization profile is at $T_{\rm meas.}=4.8$ not yet reached for $L =32$, whereas at higher temperatures this happens already for smaller $L$'s (left frame).

Scaling of the current $j \sim 1/\sqrt{L}$ or, equivalently, of diffusion constant $D \sim \sqrt{L}$, can be used to show in a non-rigorous way the spatial dependence of the magnetization. We shall use the Fourier law (\ref{eq:Four}) with a space-dependent diffusion constant $D(r)$ at site $r$. Because equation (\ref{eq:D}) tells us that the diffusion constant gets larger for larger chains, close to boundaries, local diffusion constant should become smaller. Assuming a square-root scaling we must have $D(r) \propto \sqrt{r(L-r)/L}$. This gives a differential equation
\begin{equation}
j=-\frac{{\rm const.}}{\sqrt{L}}= -\sqrt{\frac{r(L-r)}{L}} \frac{dz}{dr},
\end{equation}
where $z=\ave{\sigma_r^{\rm z}}$. Integrating the above equation with appropriate boundary conditions one immediately gets the profile $z \sim \arcsin{(2r/L-1)}$.

\subsection{Spectral properties of ${\cal L}$}
\label{sec:33}

Spectral properties of a Liouvillean superoperator ${\cal L}$, i.e., the linear operator representing the right-hand-side of the master equation (\ref{eq:Lin}), are important for several reasons. For instance, they determine the relaxation rate to NESS as well as deviations from NESS expectation values at finite times. The superoperator ${\cal L}$ is non-Hermitean and therefore has a spectrum of eigenvalues $\lambda_k, k=0,\ldots,4^L-1$, lying in a complex plane. We shall order eigenvalues $\lambda_k$ in a descending order according to their real part, starting with the largest $\lambda_0=0$. The right eigenvector $\ket{x^{\rm R}_0}$, corresponding to $\lambda_0$, is the sought-for NESS state $\rho$, symbolically $\ket{x^{\rm R}_0}=\ket{\rho}$. As a consequence of trace preservation of ${\cal L}$ the left eigenvector $\bra{x^{\rm L}_0}$ corresponding to $\lambda_0$ is on the other hand proportional to the identity operator $\sim \mathbbm{1}$, irrespective of the system, $\bra{x^{\rm L}_0}=\bra{\mathbbm{1}}$. In our system $\lambda_0$ is always nondegenerate. For simplicity we shall in this subsection discuss properties of the isotropic Heisenberg model with a one-spin bath, that is at an infinite temperature. The driving potential is weak, $\mu=0.02$, however, the values of the eigenvalues are in the linear response regime largely independent of $\mu$.

Besides $\ket{\rho}$, eigenvalues and eigenvectors that are closest to $\lambda_0$ (in real part) are also of interest. Because the dynamics governed by the Lindblad equation is contractive all real parts of eigenvalues are non-positive, ${\rm Re}(\lambda_k)\le 0$. Linear operator ${\cal L}$ is in general non-diagonalizable with the Jordan canonical form, see for instance Ref.~\cite{tomaz10} for a discussion of spectral decomposition in such case for quadratic fermionic systems. For isotropic Heisenberg model (\ref{eq:Lin}) we have found by numerical computation that for few eigenvalues with the largest real parts there are always as many linearly independent eigenvectors as is the multiplicity of the corresponding eigenvalue and therefore the Jordan form for these eigenvalues is trivial of dimension $1$. We can therefore write 
\begin{equation}
{\cal L}=\lambda_0\, \ket{x^{\rm R}_0}\bra{x^{\rm L}_0}+\lambda_1\, \ket{x^{\rm R}_1}\bra{x^{\rm L}_1}+\lambda_2\, \ket{x^{\rm R}_2}\bra{x^{\rm L}_2}+ \cdots,
\label{eq:Ldec}
\end{equation}
where left and right eigenvectors are mutually orthogonal, $\braket{x^{\rm L}_j}{x^{\rm R}_k}=\delta_{jk}$, with the standard Hilbert-Schmidt inner product $\braket{A}{B}=\tr{(A^\dagger B)}$, and we normalize left eigenvectors $\bra{x^{\rm L}_j}$. Then the value of the real part of $\lambda_1$, also called the gap of Liouvillean, determines the convergence rate with which the NESS is reached from $\rho(0)$, while the corresponding eigenvector gives the deviation of $\rho(t)$ from the NESS $\rho$. In addition, the scaling of the gap with the system size $L$ can be used to locate nonequilibrium phase transitions. For studies of this phenomenon in quantum system see Ref.~\cite{PRE11}, for classical systems see e.g.~\cite{classical}.

\begin{figure}[ht!]
\centerline{\includegraphics[width=0.7\textwidth]{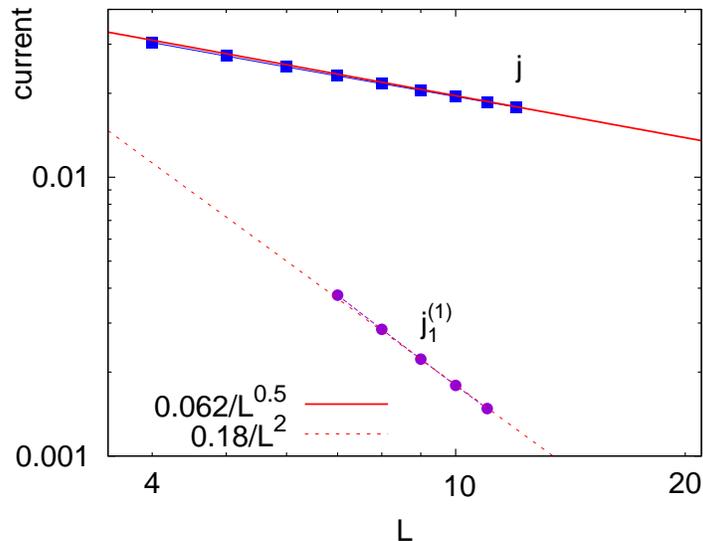}}
\caption{The current in the NESS state, $j$ (squares; the same data as in Fig.~\ref{fig:j_odn}), and the current in the 2nd eigenvector, corresponding to $\lambda_1$, at the 1st site (where it is the largest), $j^{(1)}_1=\braket{j_1}{x^{\rm R}_1}$. All using exact diagonalization and a one-spin bath (i.e., $T=\infty$).}
\label{fig:lam1}
\end{figure}

We have determined the lowest 4 eigenvalues $\lambda_{0,1,2,3}$ and their corresponding left and right eigenvectors using numerically exact diagonalization on small systems of size $L \le 12$. In addition, we determined the relaxation rate $r$ of our tDMRG solution $\rho(t)$ by fitting the convergence of magnetization at the middle of chain to its asymptotic value, $\tr{(\rho(t) \sigma^{\rm z}_{L/2})}-\ave{\sigma_{L/2}^{\rm z}}\sim \exp{(-r t)}$ (the same $r$ is also obtained by looking at the convergence of magnetization current). 
\begin{figure}[ht!]
\centerline{\includegraphics[width=0.7\textwidth]{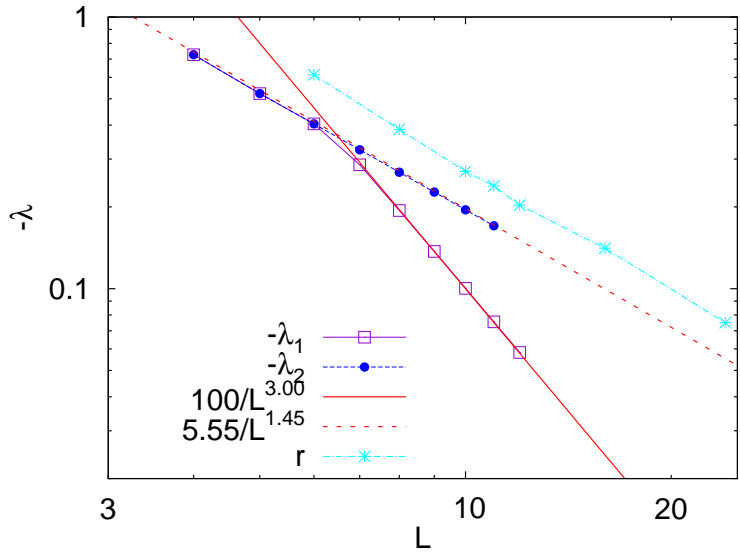}}
\caption{Gap of the superoperator ${\cal L}$, eq. (\ref{eq:Lin}), for a one-spin bath (i.e., infinite temperature) from an exact calculation (squares and circles) and relaxation rate of tDMRG simulations $r$ (crosses). The largest nontrivial eigenvalue of ${\cal L}$ scales as $\lambda_1 \sim -100/L^{3.0}$ (full line), while the 2nd one goes as $\lambda_2 \sim -5.55/L^{1.45}$ (dashed line). The scaling of tDMRG convergence rate $r$ with $L$ is the same as that of $\lambda_2$.}
\label{fig:gap}
\end{figure}
In our isotropic Heisenberg chain with one-spin bath $\lambda_1$ and $\lambda_2$ always have zero imaginary part, ${\rm Im}(\lambda_{1,2})=0$. Data in Fig.~\ref{fig:gap} shows that the gap of ${\cal L}$ decreases as $\lambda_1 \sim 1/L^3$. The same scaling with $L$ is obtained also for the XX model~\cite{iztok}. The eigenvalue $\lambda_1$ is $2\times$ degenerate for $L\le 6$, while it is nondegenerate for $L > 6$. On the other hand, $\lambda_2$ becomes $2\times$ degenerate for $L > 6$. The scaling of $\lambda_2$ for large $L$ is $\lambda_2 \sim 1/L^{1.45}$ and therefore decays with $L$ in a much slower way than $\lambda_1$. What is interesting is that the convergence rate of tDMRG simulation is not given by $\lambda_1$, but rather follows the scaling of $\lambda_2$. This is so because expectation values of current and magnetization, which are relevant observables for our discussion of transport, are very small in the eigenvector corresponding to $\lambda_1$. In fact, for $L \le 6$ they are identically zero, while for larger $L$ their values are shown in Fig.~\ref{fig:lam1}.
One can see that the contribution from $\ket{x^{\rm R}_1}$ to the magnetization current (and magnetization as well) scales as $\sim 1/L^2$ and is indeed negligible in the thermodynamic limit~\cite{footj1}. For instance, relative contribution at $L=128$ would be $(0.18/L^2)/(0.062/L^{0.5}) \approx  0.002$, which is below the precision of our tDMRG simulations. Namely, we estimate that the truncation errors in tDMRG simulations result in relative error in the current below $1\%$ at $T_{\rm meas.}=\infty$ and below $5\%$ at $T_{\rm meas.}=4.8$, both for the largest sizes shown.

\section{Conclusion}
Using extensive numerical calculations of nonequilibrium steady states close to equilibrium in chains of upto $256$ spins we have shown that the diffusion constant of magnetization in the isotropic Heisenberg model scales with the system length as $D \sim L^{0.5}$ for temperatures larger than the value of the exchange interaction. In this temperature regime the anomalous diffusion exponent of $0.5$ seems largely independent of the temperature. The spectral properties of the Liouville superoperator have also been explored, showing that that the gap scales as $\sim 1/L^3$ with the system length $L$, while the tDMRG expectation values of magnetization and current converge on a faster time increasing as $\sim L^{1.45}$.


\begin{thebibliography}{}
\bibitem{footnote} If we would have a prefactor $1/4$ in the $H$, i.e., have spin variables instead of Pauli matrices, the temperature $T$, energy density $\ave{h}$ and diffusion constant $D$ given in the paper have to be divided by $4$, whereas the magnetization current given in the paper would have to be divided by $8$.
\end{thebibliography}
\end{document}